\documentclass[12pt,superscriptaddress,showpacs,showkeys,twocolumn,prb,aps,reprint,a4paper,floatfix]{revtex4-2}

\usepackage{graphicx}
\usepackage{amsmath, amsthm, amssymb, amsfonts}
\usepackage{float}

\usepackage{natbib}
\usepackage{physics}
\usepackage[para,online,flushleft]{threeparttable} 
\usepackage{array}
\usepackage{ragged2e} 

\makeatletter
\newcommand*{\citenst}[2][]{%
  \begingroup
  \let\NAT@mbox=\mbox
  \let\@cite\NAT@citenum
  \let\NAT@space\NAT@spacechar
  \let\NAT@super@kern\relax
  \renewcommand\NAT@open{[}%
  \renewcommand\NAT@close{]}%
  \citet[#1]{#2}%
  \endgroup
}
\makeatother

\makeatletter
\newcommand*{\citenumns}[2][]{%
  \begingroup
  \let\NAT@mbox=\mbox
  \let\@cite\NAT@citenum
  \let\NAT@space\NAT@spacechar
  \let\NAT@super@kern\relax
  \renewcommand\NAT@open{[}
  \renewcommand\NAT@close{]}%
  \cite[#1]{#2}
  \endgroup
}
\makeatother
\usepackage[caption=false]{subfig}

\usepackage{lipsum}

\begin{document}
\title{Non-degenerate pumping of superconducting resonator parametric amplifier with evidence of phase-sensitive amplification}

\author{Songyuan Zhao}
\email{Author to whom any correspondence should be addressed.\\ E-mail: songyuan.zhao@physics.ox.ac.uk}
\affiliation{Clarendon Laboratory, Parks Road, Oxford OX1 3PU, United Kingdom.}
\author{S. Withington}
\affiliation{Clarendon Laboratory, Parks Road, Oxford OX1 3PU, United Kingdom.}
\author{C. N. Thomas}
\affiliation{Cavendish Laboratory, JJ Thomson Avenue, Cambridge CB3 OHE, United Kingdom.}
\date{\today}

\begin{abstract}
\noindent Superconducting resonator parametric amplifiers are potentially important components for a wide variety of fundamental physics experiments and utilitarian applications. We propose and realise an operating scheme that achieves amplification through the use of non-degenerate pumps, which addresses two key challenges in the design of parametric amplifiers: non-continuous gain across the amplification band and pump-tone removal. We have experimentally demonstrated the non-degenerate pumping scheme using a half-wave resonator amplifier based on NbN thin-film, and measured a peak gain of $26\,\mathrm{dB}$ and 3-dB bandwidth of $0.5\,\mathrm{MHz}$. The two non-degenerate pump tones were positioned $\sim10$ bandwidths above and below the frequency at which peak gain occurs. We have found the non-degenerate pumping scheme to be more stable compared to the usual degenerate pumping scheme in terms of gain drift over time, by a factor of 4. This scheme also retains the usual flexibility of NbN resonator parametric amplifiers in terms of reliable amplification in a $\sim4\,\mathrm{K}$ environment, and is suitable for cross-harmonic amplification. The use of pump tones at different frequencies allows phase-sensitive amplification when the signal tone is degenerate with the idler tone. A gain of $23\,\mathrm{dB}$ and squeezing ratio of $6\,\mathrm{dB}$ were measured. 
\end{abstract}


\maketitle

\section{Introduction}
We describe a method for operating kinetic inductance parametric amplifiers using non-degenerate pump tones, which has important practical advantages and enables phase-sensitive amplification.

Superconducting parametric amplifiers have seen significant development over recent decades due to their ability to achieve noise levels approaching the Standard Quantum Limit (SQL) of added noise, which is nearly an order of magnitude lower compared to high electron mobility transistor (HEMT) amplifiers based on semiconductors \citenumns{Eom_2012,McCulloch_2017}. Parametric amplifiers can be broadly categorised into either travelling-wave or resonator architectures. Both architectures can achieve high gain exceeding $20\,\mathrm{dB}$ with added noise close to the SQL \citenumns{Macklin_2015,Eom_2012,Tholen_2009}. Travelling-wave parametric amplifiers have very wide amplification band of several $\mathrm{GHz}$, but suffer from rapid and intense gain variations of $\sim10\mathrm{dB}$ over tens of $\mathrm{MHz}$. Resonator amplifiers (ResPAs), on the other hand, have limited amplification bandwidth (from several to hundreds of $\mathrm{MHz}$), but they have lower pump power requirements and are less prone to lithographic defects such as shorts or breaks on narrow transmission lines with micron-scale features \citenumns{Eom_2012,Shan_2016,zhao2022physics} since the transmission lines are typically much shorter for resonator amplifiers ($\lesssim1\,\mathrm{cm}$) as compared to travelling-wave amplifiers ($\gtrsim 0.5\,\mathrm{m}$). 


The ease-of-fabrication of ResPAs means tens to hundreds of amplifiers can be fabricated on a single wafer whilst maintaining extremely high yield. This technology is highly suitable for narrow-band applications that require high gain, low noise amplification, especially in the format of large arrays. These narrow-band applications range from fundamental physics, for example direct measurement of neutrino mass through cyclotron radiation emission spectroscopy \citenumns{Oblath_2020,Saakyan_2020,QTNM_collaboration_white_paper}, dark matter searches \citenumns{Axion_2024,HiddenSector_2024,Adams_2022_Snowmass}, and readout of detectors for astronomy \citenumns{HAMILTON_1980,Westig_2018}, to quantum computing and control, for example readout of qubits \citenumns{Devoret_2013,Naaman_2022}, quantum feedback \citenumns{Vijay2012_feedback}, quantum error detection \citenumns{Córcoles_2015,Riste_2015}, and measurement of quantum nanomechanical oscillators and nanobolometers \citenumns{CLeland_2002,Teufel_2011,Kokkoniemi_2019}.

Superconducting parametric amplifiers can be realised by exploiting either the Josephson junction nonlinearity, as in the case of Josephson Parametric Amplifiers (JPAs), or the kinetic inductance of superconducting thin-films, as in the case of Kinetic Inductance Travelling-Wave Parametric Amplifiers (KI-TWPAs) and Kinetic Inductance Resonator Parametric Amplifiers (KI-ResPAs). Amplifiers based on both types of nonlinearities have demonstrated added noise close to the SQL \citenumns{Yurke_1988,Malnou_2020}. In this paper, we focus on the KI-ResPAs technology, which has lower fabrication requirements (e.g. single layer coplanar waveguide with no sub-$\mu m$ features), and can be designed to have the benefit of higher saturation powers, higher maximum operating frequencies, and higher operating temperatures compared to JPAs \citenumns{Zhao_2023,zhao2024_intrinsic_separation}.

The narrow bandwidths of KI-ResPAs come with two challenges relating to their operation: Firstly, pump removal is practically challenging as the pump tone cannot be easily removed by filtering, which now requires very high-Q tunable filtering so as not to attenuate the entire amplification band inadvertently. Alternatively, the pump can be removed by splitting into two paths, and combining out-of-phase after one path is used to achieve parametric amplification. This technique also has its practical limitations as it requires delicate phase matching, tunable attenuators and phase-shifters, and additional paths in the cryostat. A circulator based pump removal scheme has recently been proposed for two-port resonator amplifiers \citenumns{zhao2024_intrinsic_separation}.  Secondly, the placement of the pump tone in the middle of the amplification band means that the amplification band cannot be used continuously in the frequency domain due to the residual of the pump tone and its noise. For experiments that require continuous frequency measurements, only half of the amplification band, i.e. a single side around the pump tone, is useful.

To address the above challenges, we propose and realise an operating scheme for KI-ResPAs using non-degenerate pump tones: two pump tones at different frequencies are used to establish the amplification operating point, which refers to the power and frequency combination of the pump tones needed to achieve high gain. The purpose of this paper is to demonstrate the non-degenerate pumping scheme for operating KI-ResPAs and, for the first time, phase-sensitive amplification using a notionally simple superconducting thin-film resonator. To enable further development and optimisation of KI-ResPAs for quantum sensing, we systematically compare the degenerate and non-degenerate operating schemes. The non-degenerate pumping scheme crucially allows the amplification band to reside between the pump frequencies. As we shall show in later sections, the non-degenerate pumping scheme results in a continuous amplification band and allows simpler pump removal. In addition, our measurements show that there is improved stability in terms of gain drift over time, due to the placement of the pump tones away from the resonance peak. The separation of the pump tones also enables phase-sensitive amplification, which occurs when the signal frequency is equal to the idler frequency. We have measured this phase-sensitive amplification, and obtained gain of $23\,\mathrm{dB}$ and squeezing ratio of $6\,\mathrm{dB}$. Finally, we demonstrate that the non-degenerate pumping scheme can be realised in a cross-harmonic mode, where the pump tones are placed in adjacent harmonics compared to the signal, and at pulse tube cooler temperature of $\sim4\,\mathrm{K}$. In total, these developments and results demonstrate that KI-ResPAs, with remarkable scalability and flexibility, has the potential to be an important technology in the future of quantum sensing.

\section{Operating Principle}
The principle of non-degenerate pumping is based on the physics of four-wave mixing. The nonlinear kinetic inductance in a superconducting thin-film has the following form, up to second order in the current $I$:
\begin{align}
L = L_0\left[1+\left(\frac{I}{I_*}\right)^2\right] \, , \label{eq:nonlinear_scale}
\end{align}
where $L$ is the inductance per unit length, $L_0$ is the inductance per unit length in the absence of inductive nonlinearity, and $I_*$ characterises the scale of the nonlinearity. $L$ and $L_0$ contains contributions from both the geometry of lines and the kinetic inductance of the conductors. As demonstrated by analysing the system using the Telegrapher's equations \citenumns{zhao2022physics}, this form of nonlinear inductance naturally leads to four-wave mixing in the input tones:
\begin{align}
\omega_{\mathrm{s}}+\omega_{\mathrm{i}} = 2\omega_{\mathrm{p}}\,, \label{eq:mixing_deg}
\end{align}
where $\omega_{\mathrm{s,i,p}}$ are the angular frequencies of the signal, idler, and pump tones respectively. In a degenerate pumping scheme, the energy from the pump tone, i.e. $2\omega_{\mathrm{p}}$, is supplied by two quanta of the pump tone at the same frequency. In a non-degenerate pumping scheme, two pump tones at different frequencies are used to apply this same pump energy. As a result, the mixing equation becomes:
\begin{align}
\omega_{\mathrm{s}}+\omega_{\mathrm{i}} = \omega_{\mathrm{p1}}+\omega_{\mathrm{p2}}\,, \label{eq:mixing_nondeg}
\end{align}
where $\omega_{\mathrm{p1}}$ and $\omega_{\mathrm{p2}}$ are the angular frequencies of the non-degenerate pump tones. In the limit where the second pump tone is supplied by a DC bias current and the first pump tone is given by $\omega_{\mathrm{s}}+\omega_{\mathrm{i}} = \omega_{\mathrm{p1}}$, the operation of the parametric amplifier behaves as governed by three-wave mixing. DC biasing is also used to achieve frequency-tunability in superconducting resonators and lower pair-breaking gaps in Kinetic Inductance Detectors \citenumns{Vissers_2015,Zhao_2020}.

The splitting of a single pump tone into separate tones at two different frequencies gives important advantages to the operation of parametric amplifiers. The first main advantage is the separation of the region of peak gain from the pump tones. The frequency region of peak gain occurs when $\omega_{\mathrm{s}}\sim\omega_{\mathrm{i}}$. Under the degenerate mixing scheme described by equation~(\ref{eq:mixing_deg}), this necessarily means that $\omega_{\mathrm{s}}\sim\omega_{\mathrm{p}}$. As a result, this high gain region is not useful experimentally due to contamination from the pump tone and its noise. Under the non-degenerate mixing scheme, as we shall show in the measurements section, the two pump tones of a resonator parametric amplifier can be placed far from the region of peak gain with $\Delta\omega_{\mathrm{p12}}>10\,\Delta\omega_{\mathrm{3dB}}$, where $\Delta\omega_{\mathrm{p12}}$ is the frequency detuning between the pump tones and $\Delta\omega_{\mathrm{3dB}}$ is the $\mathrm{3\,dB}$ bandwidth of the amplification profile.  This allows the entire continuous amplification band to be separated from the pump tones in frequency, and be useful experimentally. Further, for a resonator parametric amplifier with resonance harmonics equally-spaced in frequency, the pump tones can be placed in adjacent harmonics separated from the signal tone, thereby giving several $\mathrm{GHz}$ of frequency separation and allowing the pump tones to be removed using simple low-Q filtering. The second main advantage is the suitability for phase-sensitive amplification, which is an essential requirement for squeezing amplification and the generation of squeezed states. The former is crucial to achieving sub-SQL noise performance whilst the latter is important to a family of interferometric ultra-precision measurements. As understood by quantum analysis of parametric amplifiers, phase-sensitive amplification occurs when $\omega_{\mathrm{s}}=\omega_{\mathrm{i}}$ for a four-wave mixing parametric system. At this specific frequency, one quadrature of the signal is amplified without a quantum limit on the added noise, and one quadrature of the signal is squeezed with noise fluctuation below that of a vacuum state. As we shall show in the measurements section, by operating a superconducting resonator amplifier in non-degenerate pumping scheme and by controlling the signal tone using a phase-shifter, key characteristics of the amplified quadrature and the squeezed quadrature can be obtained whilst definitively demonstrating phase-sensitive amplification.

Non-degenerate pumping is an established technique in Optical Parametric Amplifiers (OPAs) \citenumns{McKinstrie_2002,Radic_2002} and recently proposed for superconducting Travelling-Wave Parametric Amplifiers (TWPAs) in a theoretical analysis \citenumns{Longden_2024}. The non-degenerate pumping scheme gives different advantages in travelling-wave and resonator geometries. In a TWPA system, the main benefit of non-degenerate pumping is the broadening and flattening of the amplification profile. The amplification profile is the result of the interplay between the electromagnetic structure and the parametric mixing induced by the nonlinearity. A TWPA is based on a transmission line which does not have, in theory, a finite bandwidth. The amplification profile is thus limited by the nonlinear mixing process. A TWPA system with non-degenerate pump tones thus seeks to broaden and flatten the amplification profile by controlling this nonlinear mixing process using two pump tones \citenumns{Longden_2024}. In contrast, a ResPA parametric amplifier is limited by the underlying electromagnetic structure, dictated by its resonator geometries. The resonance bandwidth, in the absence of nonlinear mixing, is of the order of $\sim10\,\mathrm{MHz}$ for the resonators tested in this study. This is much smaller than the mixing bandwidth from the kinetic inductance nonlinearity, which has been measured to be \textit{at least} several $\mathrm{GHz}$ for disordered materials such as NbTiN and NbN \citenumns{Eom_2012,Zhao_2023}. Thus a ResPA with non-degenerate pump tones has similar bandwidth as a ResPA with degenerate pump tones, which is limited by a fixed gain-bandwidth product \citenumns{Chris_2021_slow_nonlinear}. However, the useable bandwidth is significantly increased as the scheme allows the pump tones to be separated from the central region of high gain. 

In terms of implementation, the two geometries of superconducting parametric amplifiers place very different requirement on the non-degenerate pump tones. In a TWPA, the underlying basic transmission line supports tone propagation at any frequency. Without intervention, the pump tones can generate and cascade unwanted harmonics which siphon energy away and disrupt the amplification of the signal tone \citenumns{Eom_2012,Landauer_1960}. This process restricts the placement of the pump tones such that the second pump tone is on the harmonic frequency of the first pump tone, i.e. $\omega_{\mathrm{p2}}=2\omega_{\mathrm{p1}}$. In addition, higher harmonics (resulting from nonlinear mixing) are denied propagation by periodically modulating the impedance of the transmission line \citenumns{Longden_2024}. In order to implement the periodic impedance modulation, the TWPA transmission line needs to be designed for the purpose of non-degenerate pumping. This strict requirement is very different from the implementation of non-degenerate pumping in a ResPA system. In contrast with a transmission line, which allows all propagation modes except those denied by the periodic impedance modulation, a resonator denies all propagation modes except those permitted by the resonance. This removes the need to design an engineering solution to the problem of pump harmonic generation, as the higher harmonics quickly escape the underlying resonance, and do not disrupt parametric amplification. 
In the measurements section, we report measurements of non-degenerate pumping operation using a half-wave ResPA originally designed and tested for degenerate pumping \citenumns{zhao2024_intrinsic_separation}. This highlights the flexibility of ResPAs in terms of their mode of operation. 

\section{Measurements}
\subsection{Experiment Details}
\begin{figure}[!htb]
\includegraphics[width=8.6cm]{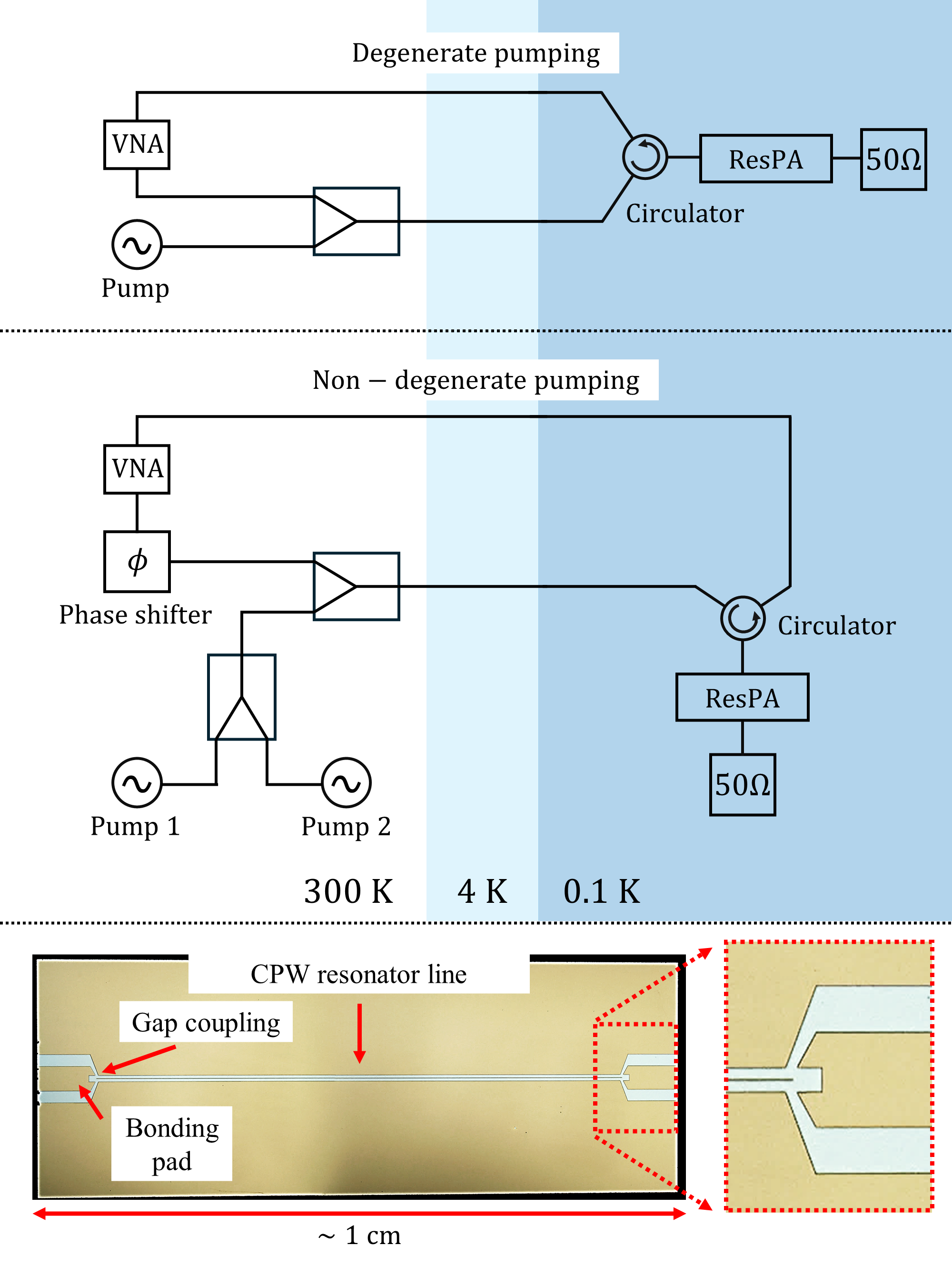}
\caption{\label{fig:00_schematic} Top subfigure: configuration of the adiabatic demagnetisation refrigerator for degenerate pumping measurements reported in \citenumns{zhao2024_intrinsic_separation}. The signal tone was supplied and measured by a vector network analyser and the pump tone was supplied by a signal generator. The tones were combined using a multiband two-way power combiner. \\ Middle subfigure: configuration of the adiabatic demagnetisation refrigerator for all non-degenerate pumping measurements. The signal tone was supplied and measured by the primary vector network analyser. The pump tones were supplied by a signal generator and a secondary vector network analyser in continuous wave mode. The tones were combined using multiband two-way power combiners. \\ Bottom subfigure: photograph of a half-wave resonator measured in this study. The resonator consisted of a length of coplanar waveguide capacitively coupled in series between two contact pads \citenumns{zhao2024_intrinsic_separation}. The length over which the coupling pad overlapped with the resonator line determined the coupling quality factor.}
\end{figure}

Details on fabrication and packaging of the particular resonator device used in this study can be found in our previous manuscript \citenumns{zhao2024_intrinsic_separation}, which also includes measurements of the resonance without pump tones and an analysis on the bifurcation characteristics of the resonator. We highlight the following key properties: The resonator consists of a single layer of thin-film NbN half-wave coplanar waveguide with a superconducting transition temperature of $\sim10\,\mathrm{K}$. $\sim25$ devices were deposited on a single 2-inch Si wafer per fabrication run. We have tested $\sim10$ NbN devices to date, and all devices produced high maximum gain exceeding $20\,\mathrm{dB}$, highlighting the reproducibility of the technology. 

The measurement setup used for the degenerate pumping measurements reported in \citenumns{zhao2024_intrinsic_separation} is shown in the top subfigure of figure~\ref{fig:00_schematic}. In the present study, the measurement system was reconfigured to enable non-degenerate pumping and phase-sensitive amplification, as shown in the middle subfigure of figure~\ref{fig:00_schematic}: the output of a signal generator and a vector network analyser (VNA) in continuous wave mode were combined using a multiband two-way power combiner before injection into the pump port; the signal tone from a separate VNA passed through an analogue phase-shifter before injection into the signal port. The phase-shifter was kept constant for all measurements unless otherwise stated. All measurements were performed in an Adiabatic Demagnetisation Refrigerator (ADR). The base temperature of the ADR is $0.1\,\mathrm{K}$; with the ADR switched off, the cryostat could be maintained at $3.2\,\mathrm{K}$ using the pulse tube cooler. The measurement temperature was $0.1\,\mathrm{K}$ unless explicitly stated to be $3.2\,\mathrm{K}$. HEMT amplifiers were not used in the signal chain as added noise measurement was not an aim of this study. 

\subsection{Experiment Results}
\begin{figure}[!htb]
\includegraphics[width=8.6cm]{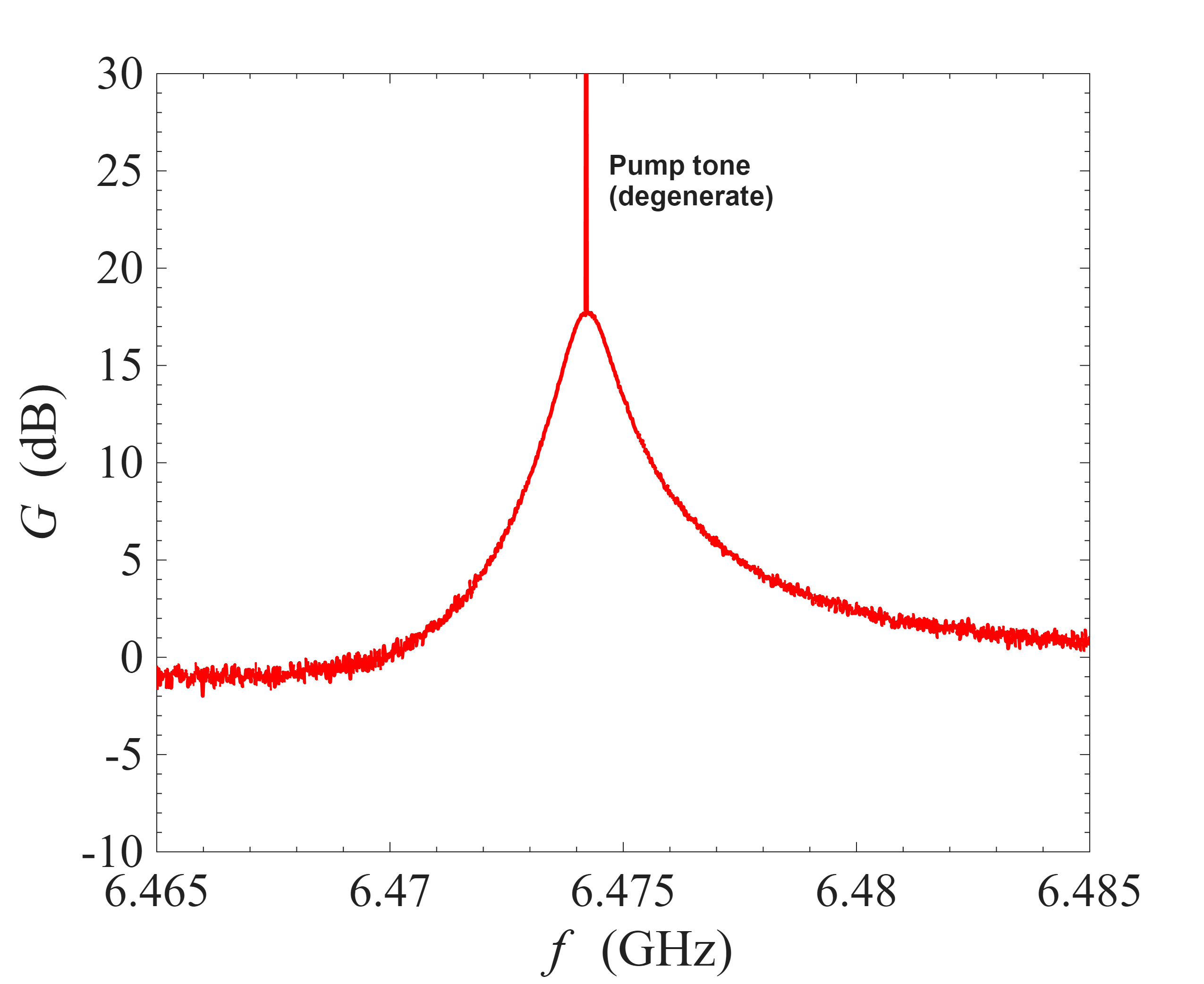}
\caption{\label{fig:6_dengenerate_pump} Gain measurement of a NbN resonator amplifier operated under degenerate pumping scheme at $0.1\,\mathrm{K}$. The frequency of the pump tones is $6.474\,\mathrm{GHz}$.}
\end{figure}

In the first instance, we operated the ResPA device under degenerate pumping, i.e. using a single pump tone, to establish a baseline behaviour for comparison against the proposed non-degenerate pumping scheme. Figure~\ref{fig:6_dengenerate_pump} shows the amplification profile from degenerate pumping using the same readout system and the same cryogenic environment of $0.1\,\mathrm{K}$. The sharp spike at $\sim6.474\,\mathrm{GHz}$ is the pump tone. As seen in the figure, the presence of the pump tone at the centre of the amplification profile interrupted the region of maximum gain, and this effect is strongly dependent on the pump removal method employed: for example, analogue filtering \citenumns{Dai_2025_JPA_filtering}, interferometric pump removal \citenumns{Eom_2012}, pump-signal separation using a circulator \citenumns{zhao2024_intrinsic_separation}, digital post-processing, or a combination of the aforementioned techniques. In addition, frequencies very close to the pump frequency will be contaminated by the phase noise of the pump (typical commercial signal generators have phase noise around $-135\,\mathrm{dBc/Hz}$ at $10\,\mathrm{GHz}$ and offset frequency of $10\,\mathrm{kHz}$ \citenumns{anritsu_SigGen}).

\begin{figure}[!htb]
\includegraphics[width=8.6cm]{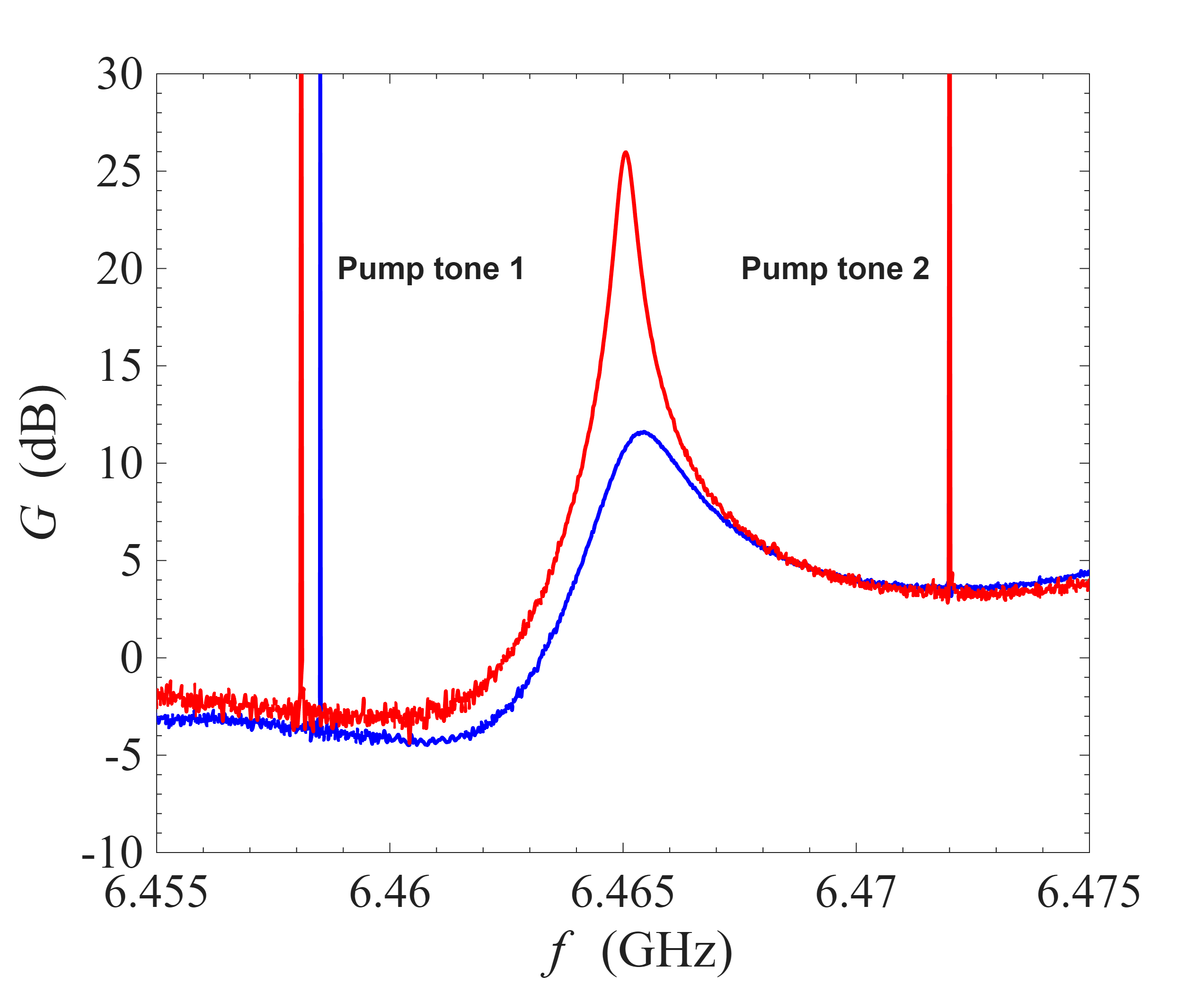}
\caption{\label{fig:1_2_gain_compiled} Gain measurement of a NbN resonator amplifier operated under non-degenerate pumping scheme at $0.1\,\mathrm{K}$. The pump tones are indicated by the sharp spikes in the figure. Red line: the frequencies of the pump tones are $6.4581\,\mathrm{GHz}$ and $6.4720\,\mathrm{GHz}$; blue line: the frequencies of the pump tones are $6.4585\,\mathrm{GHz}$ and $6.4720\,\mathrm{GHz}$.}
\end{figure}

\begin{figure*}
\includegraphics[width=16cm]{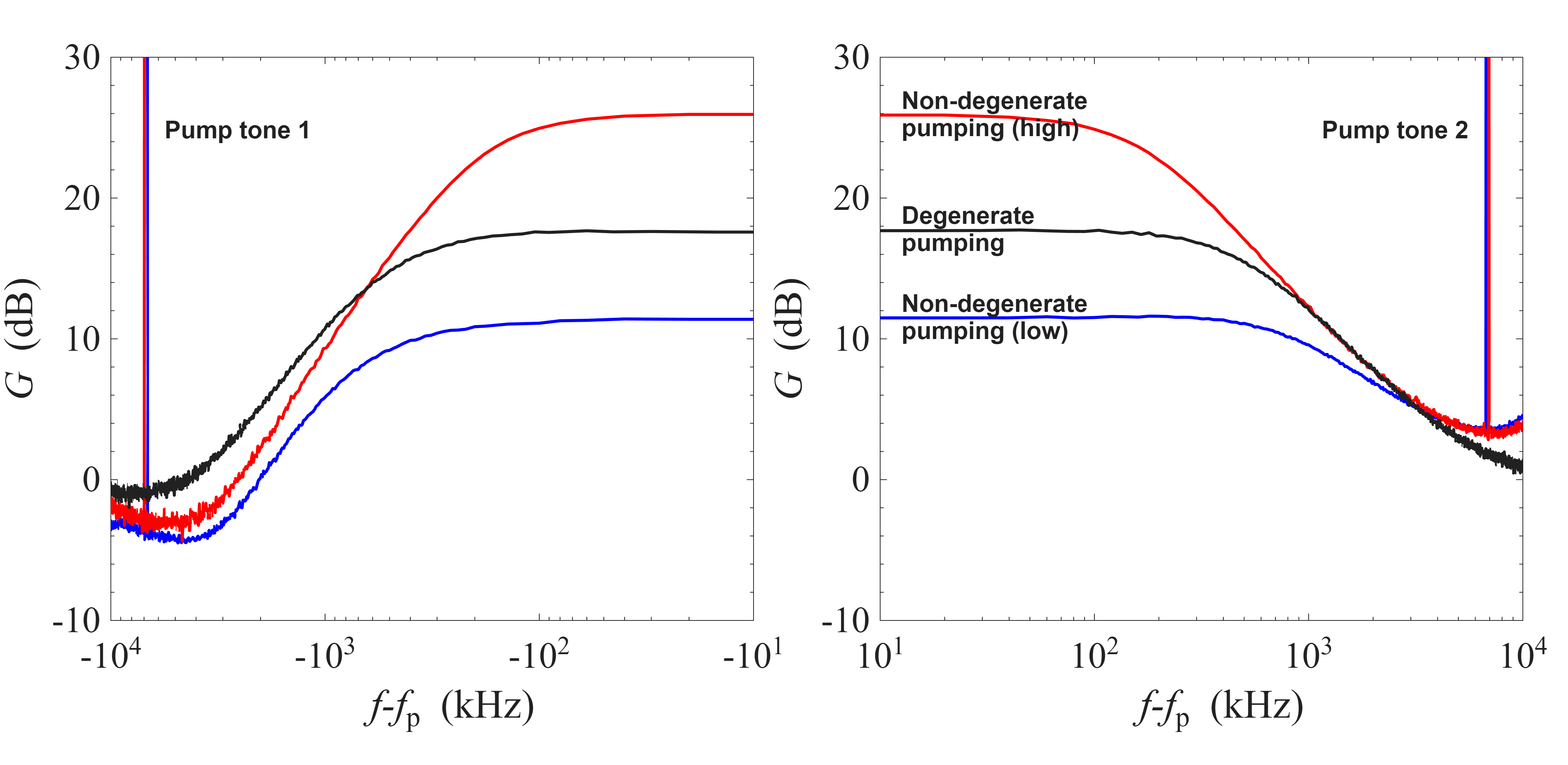}
\caption{\label{fig:14_logarithmic_compiled} Gain measurement of a NbN resonator amplifier at $0.1\,\mathrm{K}$ against a logarithmic frequency scale. Red line: non-degenerate pumping at high gain; black line: degenerate pumping at moderate gain; blue line: non-degenerate pumping at low gain.}
\end{figure*}

In contrast, figure~\ref{fig:1_2_gain_compiled} shows the amplification profile of our half-wave resonator when operated using the non-degenerate pumping scheme at $0.1\,\mathrm{K}$. The sharp spikes at $\sim6.458\,\mathrm{GHz}$ and $\sim6.472\,\mathrm{GHz}$ are the pump tones appearing in the signal sweep. As seen in the figure, one pump was placed at $\sim6.472\,\mathrm{GHz}$ while a second pump was placed at $\sim6.458\,\mathrm{GHz}$. Depending on the precise placement of the second pump, the maximum gain takes on different values. As seen in the figure, by adjusting the second pump from $6.4585\,\mathrm{GHz}$ (blue line) to $6.4581\,\mathrm{GHz}$ (red line), the peak power gain increases from $10\,\mathrm{dB}$ to $>20\,\mathrm{dB}$. This allows the amplifier to be flexible in responding to experimental needs: whilst some experiments require maximum gain, other experiments might benefit from a lower gain but a wider bandwidth around the peak gain. Figure~\ref{fig:1_2_gain_compiled} also illustrates one of the main advantages of non-degenerate pumping: as seen in the figure, both pump tones are well away from the $\mathrm{1\,dB}$ bandwidth of the amplification profile, or even the $\mathrm{3\,dB}$ bandwidth. This allows the pump tones to be removed using simple low-Q filtering techniques, and prevents the contamination of the amplification band by the noise from the pump tones.

To compare the different modes of operating a ResPA, we have plotted the three amplification profiles from the ResPA, i.e. non-degenerate high gain, non-degenerate low gain, and degenerate high gain, on a logarithmic frequency scale. For degenerate pumping, $f_\mathrm{p}$ refers to the pump frequency; for non-degenerate pumping, $f_\mathrm{p}$ refers to the mean of the pump frequencies. As figure~\ref{fig:14_logarithmic_compiled} shows, in the degenerate pumping scheme, the 1-dB bandwidth lies $0.2\,\mathrm{MHz}$ away from the pump tone, whereas in the non-degenerate scheme it is $7\,\mathrm{MHz}$ from the nearest pump tone. For a typical commercially available signal generator operating at $6\,\mathrm{GHz}$ \citenumns{keysight_sigGen}, this corresponds to phase noise levels of $-135\,\mathrm{dBc/Hz}$ and $-150\,\mathrm{dBc/Hz}$, respectively. The non-degenerate scheme therefore offers a significant noise advantage by distancing the high-gain region from the pump tones in terms of frequency.

As seen in figure~\ref{fig:14_logarithmic_compiled}, the rising/falling edges of the three amplification profiles align approximately with each other. This suggests that despite the different modes of operation, the amplification profiles are governed by a similar single-pole roll-off behaviour and have a similar gain-bandwidth product. This happens because, in a kinetic inductance ResPA, the bandwidth is limited by the underlying electromagnetic behaviour of the resonator instead of the mixing bandwidth of the nonlinearity. Despite the different modes of operation, the underlying resonator is the same and its resonance sets a baseline of gain-bandwidth product which is approximately given by $g\times\Delta\omega_{g}=1\times\Delta\omega_{1}$, where $g$ is the magnitude of the amplitude gain, $\Delta\omega_{g}$ is the bandwidth when the peak amplitude gain is $g$, and $\Delta\omega_{1}$ is the bandwidth in the absence of pump at unity gain, which is also the bandwidth of the underlying resonance \citenumns{Zhao_2023,Thomas_2022}.

\begin{figure}
\includegraphics[width=8.6cm]{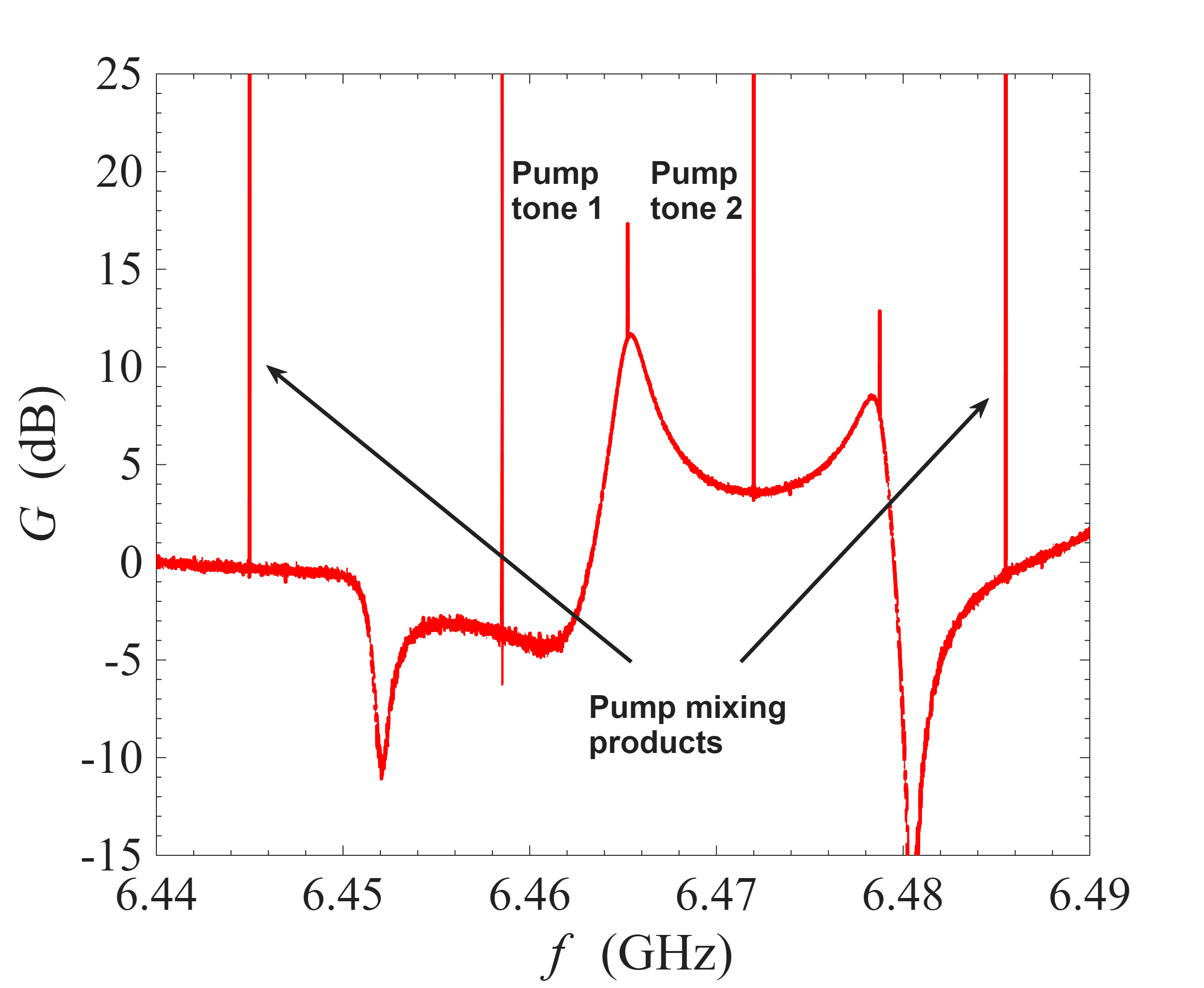}
\caption{\label{fig:3_gain_wide} Gain measurement of a NbN resonator amplifier operated under non-degenerate pumping scheme with a wide frequency scale at $0.1\,\mathrm{K}$. The frequencies of the pump tones are $6.4585\,\mathrm{GHz}$ and $6.4720\,\mathrm{GHz}$. Sharp spikes at $6.4450\,\mathrm{GHz}$ and $6.4855\,\mathrm{GHz}$ are the result of inter-mixing between the pump tones through the underlying nonlinearity. Signal spikes of $6\,\mathrm{dB}$ in height occurs at $6.465\,\mathrm{GHz}$ and $6.479\,\mathrm{GHz}$ as a result of phase-sensitive amplification.}
\end{figure}

\begin{figure*}
\includegraphics[width=16cm]{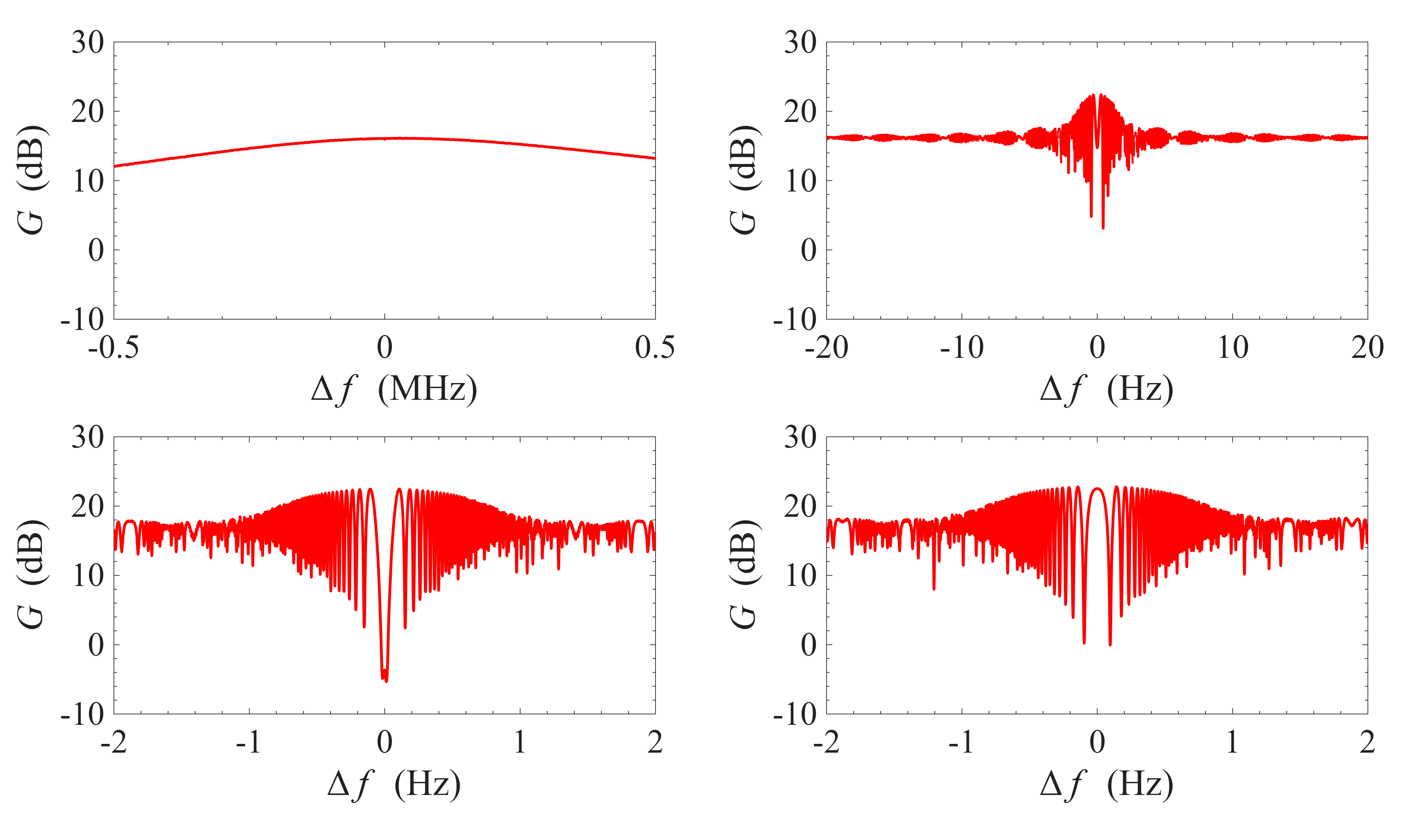}
\caption{\label{fig:8_phase_sensitivity} Gain measurement around the signal-idler degenerate frequency at $0.1\,\mathrm{K}$. Top subfigures: frequency span is progressively zoomed-in around the degenerate frequency; bottom subfigures: the phase-shifter is adjusted such that, at the degenerate frequency, the signal is maximally squeezed on the left subfigure, and maximally amplified on the right subfigure.}
\end{figure*}

Figure~\ref{fig:3_gain_wide} shows an amplification profile from non-degenerate pumping on a wide frequency axis. The frequencies of the two injected pump tones are $6.4585\,\mathrm{GHz}$ and $6.4720\,\mathrm{GHz}$. As seen in the figure, there are two types of power spikes in this figure: the pump spikes occur at $6.4450\,\mathrm{GHz}$, $6.4585\,\mathrm{GHz}$, $6.4720\,\mathrm{GHz}$, and $6.4855\,\mathrm{GHz}$ whilst the signal $6\,\mathrm{dB}$ spikes occur at $6.465\,\mathrm{GHz}$, and $6.479\,\mathrm{GHz}$. We explain the signal $6\,\mathrm{dB}$ spikes at the discussion around figure~\ref{fig:8_phase_sensitivity}, as it is a characteristic behaviour of phase-sensitive amplification against a background of phase-insensitive amplification. The pump spikes are the result of nonlinear intermixing between the two strong pump tones. Due to the nature of four-wave mixing, the pump mixing products occur at $\omega_{\mathrm{p1}}+n\Delta\omega_{\mathrm{p12}}$, where $\Delta\omega_{\mathrm{p12}}=\omega_{\mathrm{p1}}-\omega_{\mathrm{p2}}$ and $n$ is an integer. At high orders of mixing, i.e. $|n|\gg1$, the pump mixing products decrease quickly in amplitude as their propagation frequencies are no longer supported by the bandwidth of the resonance. Amplification is not supported at the region between $6.4450\,\mathrm{GHz}$ and $6.4585\,\mathrm{GHz}$ because this region does not overlap significantly with the underlying resonance. Interestingly, the region between $6.4720\,\mathrm{GHz}$ and $6.4855\,\mathrm{GHz}$ does support amplification even though the $6.4855\,\mathrm{GHz}$ tone is generated by the inter-mixing between the pump tones. The resultant peak gain is lower compared to the primary amplification region between $6.4585\,\mathrm{GHz}$ and $6.4720\,\mathrm{GHz}$, i.e. between the injected pump tones.

\begin{figure}[!htb]
\includegraphics[width=8.6cm]{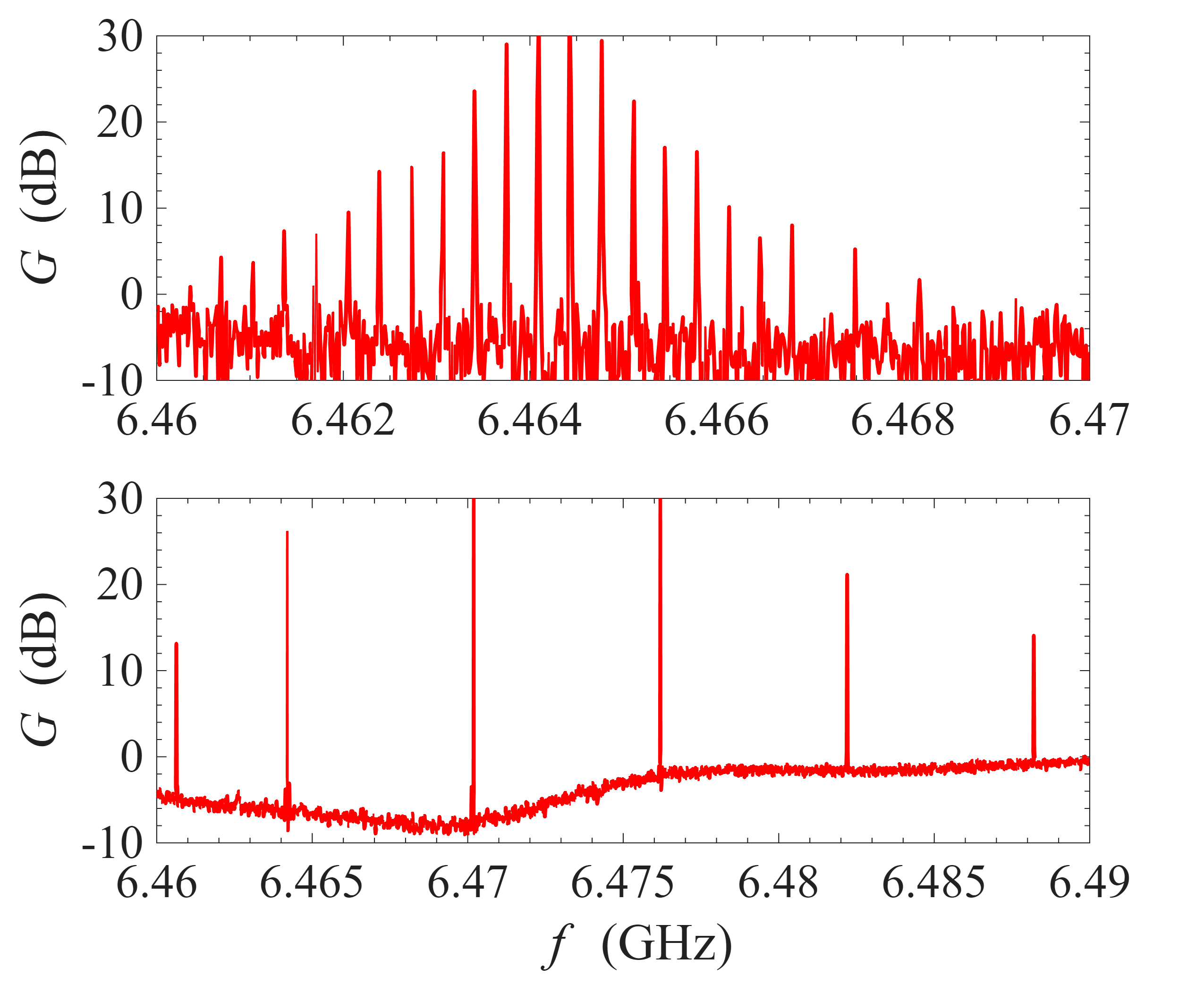}
\caption{\label{fig:4_close_missing} Top subfigure: measurement of amplifier behaviour when two pump tones are placed close in frequency at $6.4641\,\mathrm{GHz}$ and $6.46443\,\mathrm{GHz}$. The other spikes are inter-mixing products between the pump tones. No signal gain is observed. Bottom subfigure: measurement of amplifier behaviour when two pump tones are placed at $6.4702\,\mathrm{GHz}$ and $6.4762\,\mathrm{GHz}$. The pump tones miss the operating point of the underlying resonator amplifier, and as a result no signal gain is observed.}
\end{figure}

One might be tempted to smoothly transition between the two pumping schemes by splitting the single pump in the degenerate scheme into two pump tones close in frequency. The top subfigure of figure~\ref{fig:4_close_missing} shows a demonstrative result from such an attempt with $\Delta\omega_{\mathrm{p12}}=0.33\,\mathrm{MHz}$. As seen in the figure, a large number of pump mixing products were generated and there was no amplification in the signal tone. This can be understood in terms of pump tones cascading unwanted mixing products which siphons energy away and disrupts the amplification of the signal tone \citenumns{Eom_2012,Landauer_1960}.

Theoretical analysis of ResPAs shows that high-gain operation requires carefully chosen combinations of pump power and frequency: strong pump tones do not lead to amplification unless their frequencies closely match the required operating point \citenumns{Thomas_2022}. The bottom subfigure of figure~\ref{fig:4_close_missing} illustrates two pump tones that are misaligned with the operating point, resulting in no signal amplification. This restriction on the operating point is a key feature of ResPA and is also important to degenerate pumping \citenumns{Zhao_2023,zhao2024_intrinsic_separation}.

\begin{figure}[!htb]
\includegraphics[width=8.6cm]{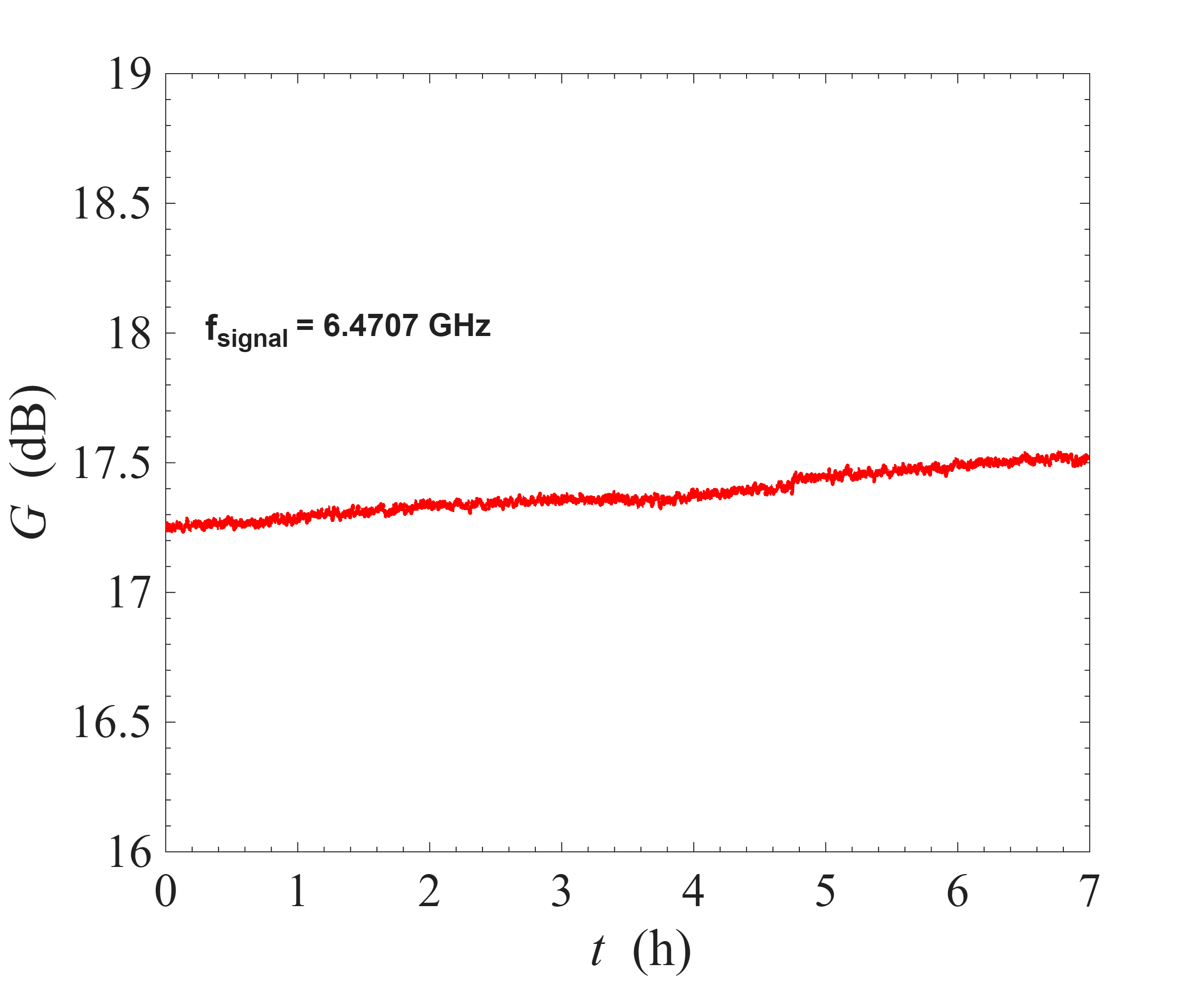}
\caption{\label{fig:7_stability_cold} Measurement of gain stability of a NbN resonator amplifier operated under non-degenerate pumping scheme at $0.1\,\mathrm{K}$. The signal frequency is fixed at $6.4707\,\mathrm{GHz}$, where the gain is close to $17\,\mathrm{dB}$, and the signal gain measured continuously as a function of time for seven hours. The amount of drift in gain is approximately $0.04\,\mathrm{dB/h}$.}
\end{figure}

\begin{figure}[!htb]
\includegraphics[width=8.6cm]{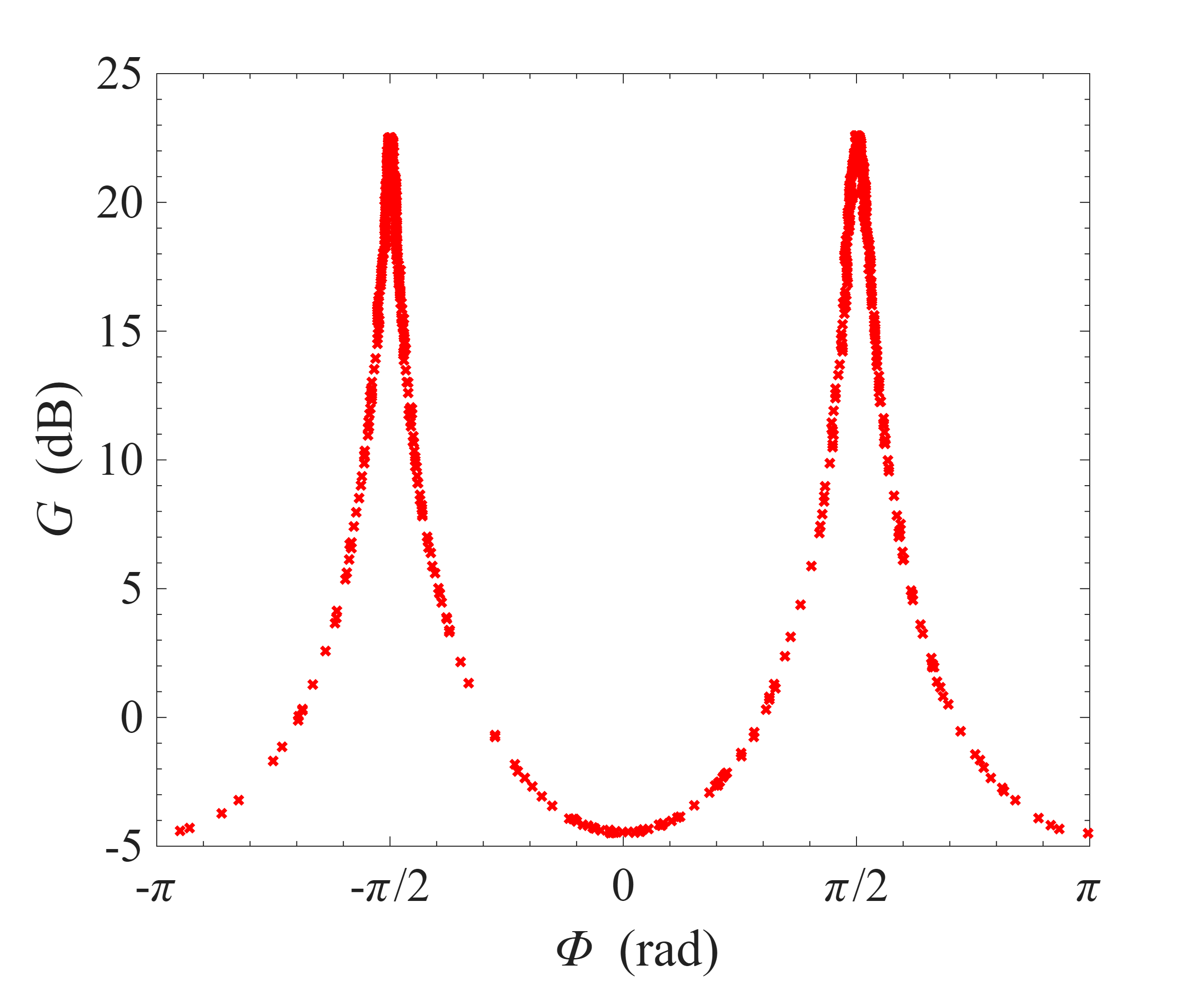}
\caption{\label{fig:9_phase_sensitivity_parametric} Gain measurement at the signal-idler degenerate frequency as a function of the output phase, which is controlled by varying the input phase using an analogue phase-shifter. Temperature of the amplifier is maintained at $0.1\,\mathrm{K}$ for this measurement.}
\end{figure}
In order to investigate the characteristic stability of the non-degenerate pumping scheme over timescale of several hours, we have fixed the signal frequency at $6.4707\,\mathrm{GHz}$ where the gain was $\sim17\,\mathrm{dB}$ and monitored the gain continuously over seven hours. As shown in figure~\ref{fig:7_stability_cold}, the gain was stable and drifted by $0.25\,\mathrm{dB}$ over seven hours, translating to a slow drift of $0.04\,\mathrm{dB/h}$. We emphasise that the very small drift has been obtained without any experimental effort at stabilising the gain. In comparison with gain drift of the same device under degenerate pumping \citenumns{zhao2024_intrinsic_separation}, which had a drift of $0.15\,\mathrm{dB/h}$, the non-degenerate pumping scheme was remarkably stable. This improvement in stability could be explained by the fact that degenerate pumping requires the pump tone to be positioned near the bifurcation point of the resonator where the transmission properties vary rapidly with frequency \citenumns{Thomas_2020,Thomas_2022,zhao2024_intrinsic_separation}. As a result, any relative drift between the pump tone and the resonance is reinforced by the proximity to the bifurcation point. Non-degenerate pumping, on the other hand, places pump tones far from the bifurcation point, thereby improving the stability of the system. The measured gain drift of $0.04\,\mathrm{dB/h}$ is comparable to that of Josephson junction-based parametric devices \citenumns{Josephson_gain_drift,Josephson_Chalmers}, which have benefited from decades of research, development, and optimisation. Josephson junction-based devices often require monitoring or calibration throughout their operation to mitigate instabilities and to establish the required accuracy \citenumns{Josephson_gain_drift,Josephson_drift_calibration}. In general, gain drift and other gain instability phenomena can significantly impact real‑world device performance, and a dedicated future study using techniques such as Allan variance will be highly valuable for the development of the ResPA technology.

As mentioned in previous sections, one important advantage of non-degenerate pumping is the possibility of the phase-sensitive amplification where $\omega_{\mathrm{s}}=\omega_{\mathrm{i}}=(\omega_{\mathrm{p1}}+\omega_{\mathrm{p2}})/2$, i.e. the signal frequency is degenerate with the idler frequency. This degeneracy condition imposes a very strict restriction on the signal frequency. As a result, features of phase-sensitive amplification cannot be observed in figure~\ref{fig:1_2_gain_compiled} where the frequency range is wide and the sampling is insufficiently fine. In figure~\ref{fig:8_phase_sensitivity}, we have progressively narrowed the range of frequency sweeps whilst keeping the number of sampling frequencies constant at 2001. Under this measurement condition, the features of phase-sensitive amplification only revealed itself significantly over a narrow window of $\pm 1\,\mathrm{Hz}$ around the degenerate frequency $\omega_{\mathrm{d}}=(\omega_{\mathrm{p1}}+\omega_{\mathrm{p2}})/2$. In the bottom two plots, the phase-shifter was adjusted such that the bottom left plot corresponds approximately to maximum signal squeezing and the bottom right plot corresponds approximately to maximum signal amplification. 

\begin{figure*}
\includegraphics[width=16cm]{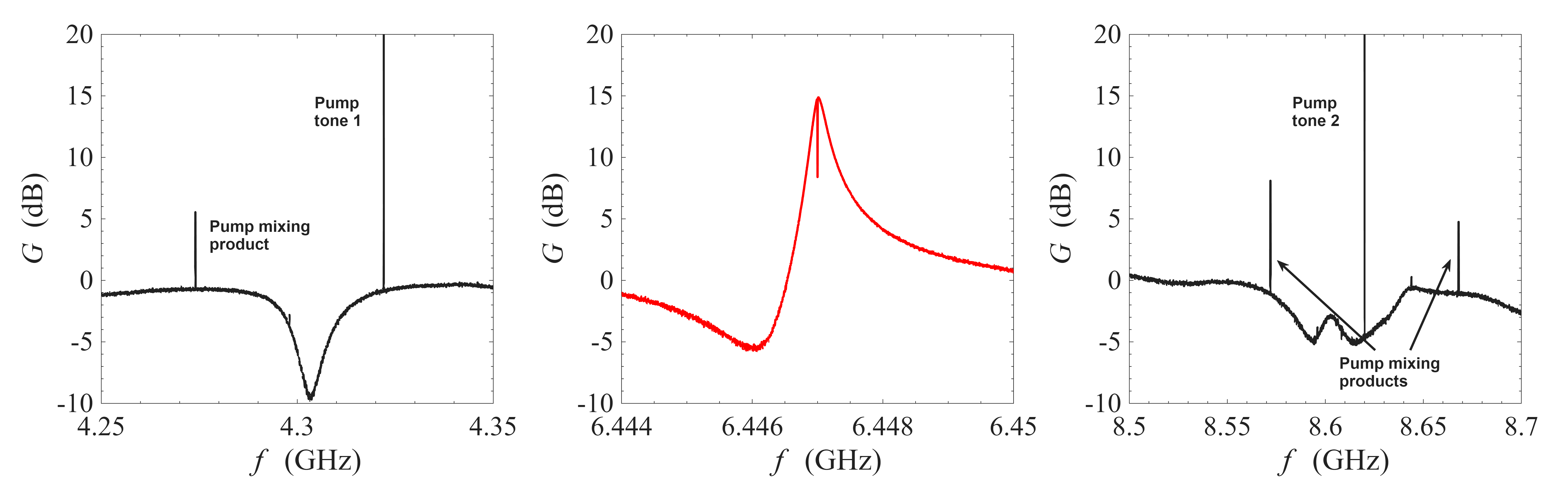}
\caption{\label{fig:10_cross_harmonic_compiled} Gain measurement of a NbN resonator amplifier operated under cross-harmonic non-degenerate pumping scheme at $0.1\,\mathrm{K}$. The pump tones are placed in adjacent harmonics around the signal harmonic, and are indicated by the sharp spikes in the left and right subfigures. The frequencies of the pump tones are $4.322\,\mathrm{GHz}$ and $8.620\,\mathrm{GHz}$, which gives a signal-idler degenerate frequency of $6.471\,\mathrm{GHz}$. This corresponds to the sharp, dipping spike in the middle figure, whose transmission amplitude can be controlled by adjusting the variable phase-shifter.}
\end{figure*}

The phase sensitivity of the amplification process can be better appreciated by plotting the gain at the signal-idler degenerate frequency against the output phase, which is varied slowly by manually controlling the phase-shifter. As seen in figure~\ref{fig:9_phase_sensitivity_parametric}, the gain varies periodically with the output phase with a period of $\pi$ instead of $2\pi$, i.e. the amplified mode and the squeezed mode differ by a quadrature phase. This is a key characteristic of squeezing amplification \citenumns{Walls1983_squeezing,Slusher_1985,Zhao_2021}. Compared to figure~\ref{fig:8_phase_sensitivity}, the maximum gain at this signal-idler degenerate frequency at $23\,\mathrm{dB}$ is exactly $6\,\mathrm{dB}$ higher compared to the gain plateau surrounding the degenerate frequency at $17\,\mathrm{dB}$. This is because: at non-degenerate frequencies, the signal power is given simply by $P_\mathrm{s}=|V_\mathrm{s}|^2/Z_0$, where $Z_0$ is the system impedance and $V_\mathrm{s}$ is the signal voltage amplitude; at degenerate frequencies, however, the signal frequency coincides with the idler frequency, and the total power is given by $P_\mathrm{s}=|V_\mathrm{s}+V_\mathrm{i}|^2/Z_0 \sim 4 |V_\mathrm{s}|^2/Z_0$, where $V_\mathrm{i}$ is the idler voltage amplitude and the approximation $V_\mathrm{i}\sim V_\mathrm{s}$ is valid at high gain \citenumns{Chaudhuri_2015,zhao2022physics}. This doubling of amplitude leads to a factor of 4 in power, thus resulting in a peak of $6\,\mathrm{dB}$ above the non-degenerate gain plateau. We also note that the maximum amount of squeezing is $5\,\mathrm{dB}$ and is much smaller than the maximum gain, which is $23\,\mathrm{dB}$. This is because, for an amplifier operating in reflection mode, the signal includes contributions from reflections off every connector / interface. At high gain, these contributions are negligible. However, at high squeezing ratios, where the signal from the parametric amplifier is small, these contributions become significant and limit the squeezing ratio. If the interfaces are sufficiently matched or the device is operated in transmission mode, the maximum squeezing ratio should approach the maximum gain.

As described in the experimental details section, the resonator parametric amplifier in this study is constructed from a length of transmission line terminated with two capacitive stubs. This naturally allows the device to have strong resonance modes that are evenly spaced in frequency. A practically-advantageous way of operating these amplifiers is by pumping on the two adjacent resonance modes around a central amplification resonance. This allows GHz-level of frequency separation between the amplification band and the pump tones, thereby preventing contamination from pump noise and enabling easier pump removal through low-Q filtering. Figure~\ref{fig:10_cross_harmonic_compiled} shows realisation of this cross-harmonic non-degenerate pumping scheme using the same amplifier as in previous measurements: pump tones were placed on the $4.3\,\mathrm{GHz}$ harmonic and the $8.6\,\mathrm{GHz}$ harmonic to enable amplification in the $6.45\,\mathrm{GHz}$ harmonic with peak gain of $15\,\mathrm{dB}$. As observed in the left and right subfigures, low-power sidebands are generated in the process: the spacing between the two strong tones (with power $>3\,\mathrm{dB}$ relative to the sweep tone) on the left subfigure, and the three strong tones in the right subfigure is $0.048\,\mathrm{GHz}$, which is twice the detuning between the pump tones, i.e. $2\times4.322\,\mathrm{GHz}-8.62\,\mathrm{GHz}=0.024\,\mathrm{GHz}$. This indicates that the sidebands are products from the inter-mixing between the pump tones. As they are much lower in power compared to the principal pump tones (by more than $50\,\mathrm{dB}$), they do not affect the amplification process.

\begin{figure}[!htb]
\includegraphics[width=8.6cm]{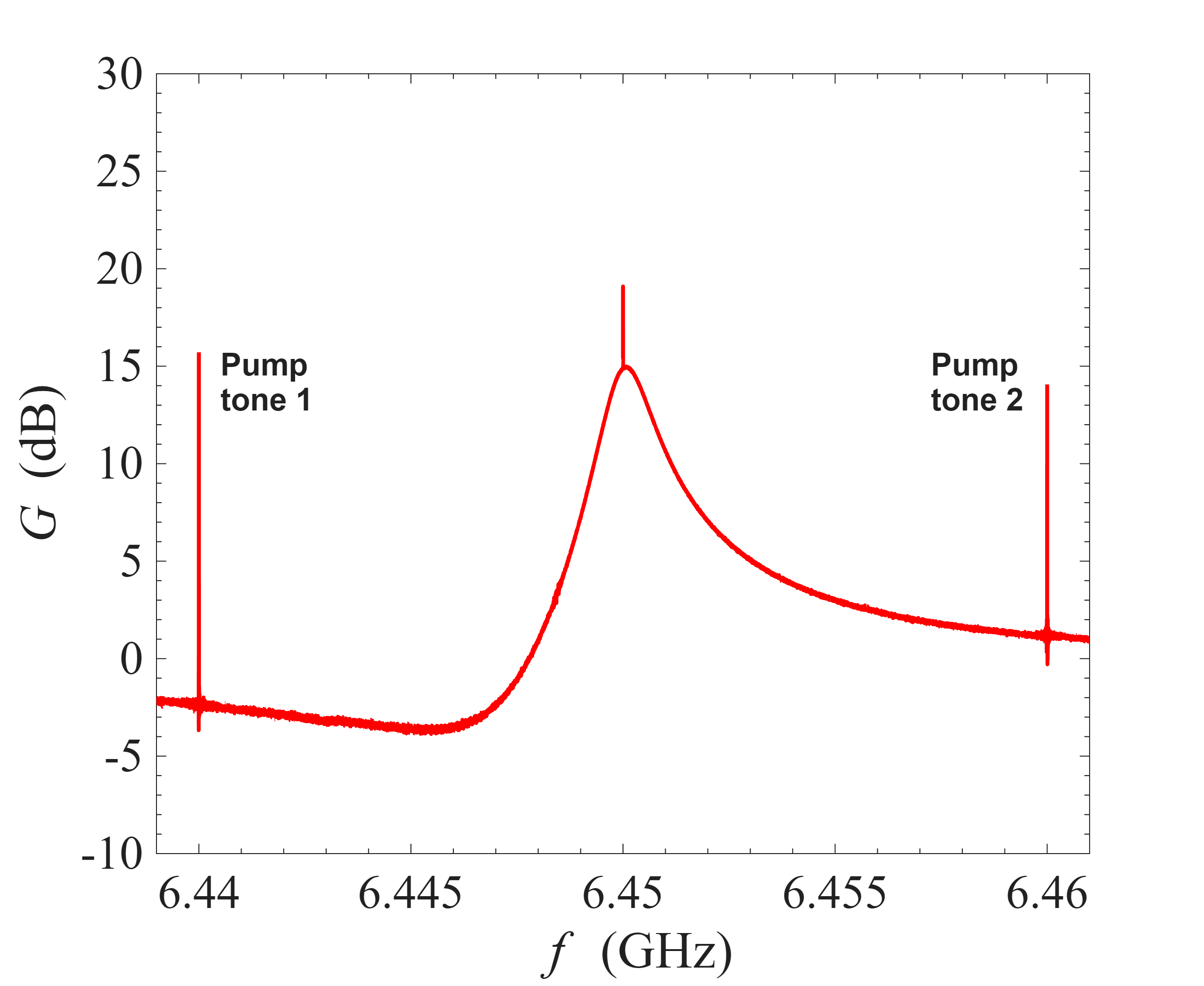}
\caption{\label{fig:11_4K_gain} Gain measurement of a NbN resonator amplifier operated under non-degenerate pumping scheme at $3.2\,\mathrm{K}$. The pump tones are indicated by the sharp spikes in the figure at $6.44\,\mathrm{GHz}$ and $6.46\,\mathrm{GHz}$.}
\end{figure}
Thus far in this section, all measurements were performed at $0.1\,\mathrm{K}$. An important application of the ResPA technology is amplification at pulse tube cooler temperatures, i.e. $\sim 4\,\mathrm{K}$, which would significantly reduce the expense and complexity of the cryostat system needed for operating a ResPA. Such operation is feasible for our ResPA because it is based on high-resistivity NbN thin-films, which have superconducting transition temperatures of $\sim10\,\mathrm{K}$. Figure~\ref{fig:11_4K_gain} shows the amplification of the same ResPA at $3.2\,\mathrm{K}$ using the non-degenerate pumping scheme. As seen in the figure, high gain of $15\,\mathrm{dB}$ has been achieved with two pumps at $6.44\,\mathrm{GHz}$ and $6.46\,\mathrm{GHz}$. 

\begin{figure}[!htb]
\includegraphics[width=8.6cm]{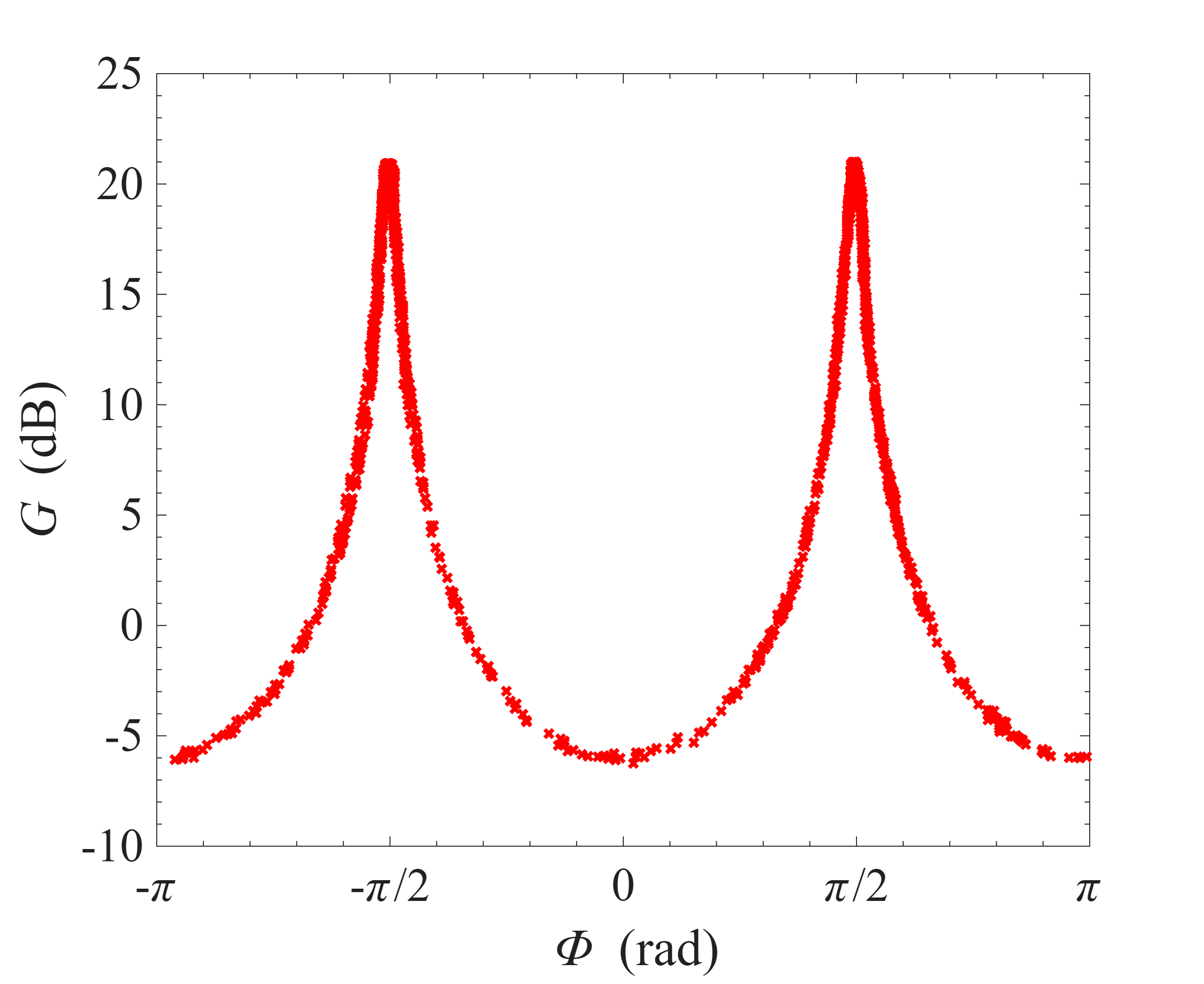}
\caption{\label{fig:12_4K_parametric} Gain measurement at the signal-idler degenerate frequency as a function of the output phase, which is controlled by varying the input phase using an analogue phase-shifter. Temperature of the amplifier is maintained at $3.2\,\mathrm{K}$ for this measurement.}
\end{figure}
Similar to the operation at $0.1\,\mathrm{K}$, at the signal-idler degenerate frequency of $6.45\,\mathrm{GHz}$, the gain/squeezing can be controlled by altering the signal phase. This phase dependence is shown in figure~\ref{fig:12_4K_parametric}. Overall, the behaviour of the device at $3.2\,\mathrm{K}$ is qualitatively identical to that at $0.1\,\mathrm{K}$, indicating that the increase in operating temperature did not adversely affect device operation.

\begin{figure}[!htb]
\includegraphics[width=8.6cm]{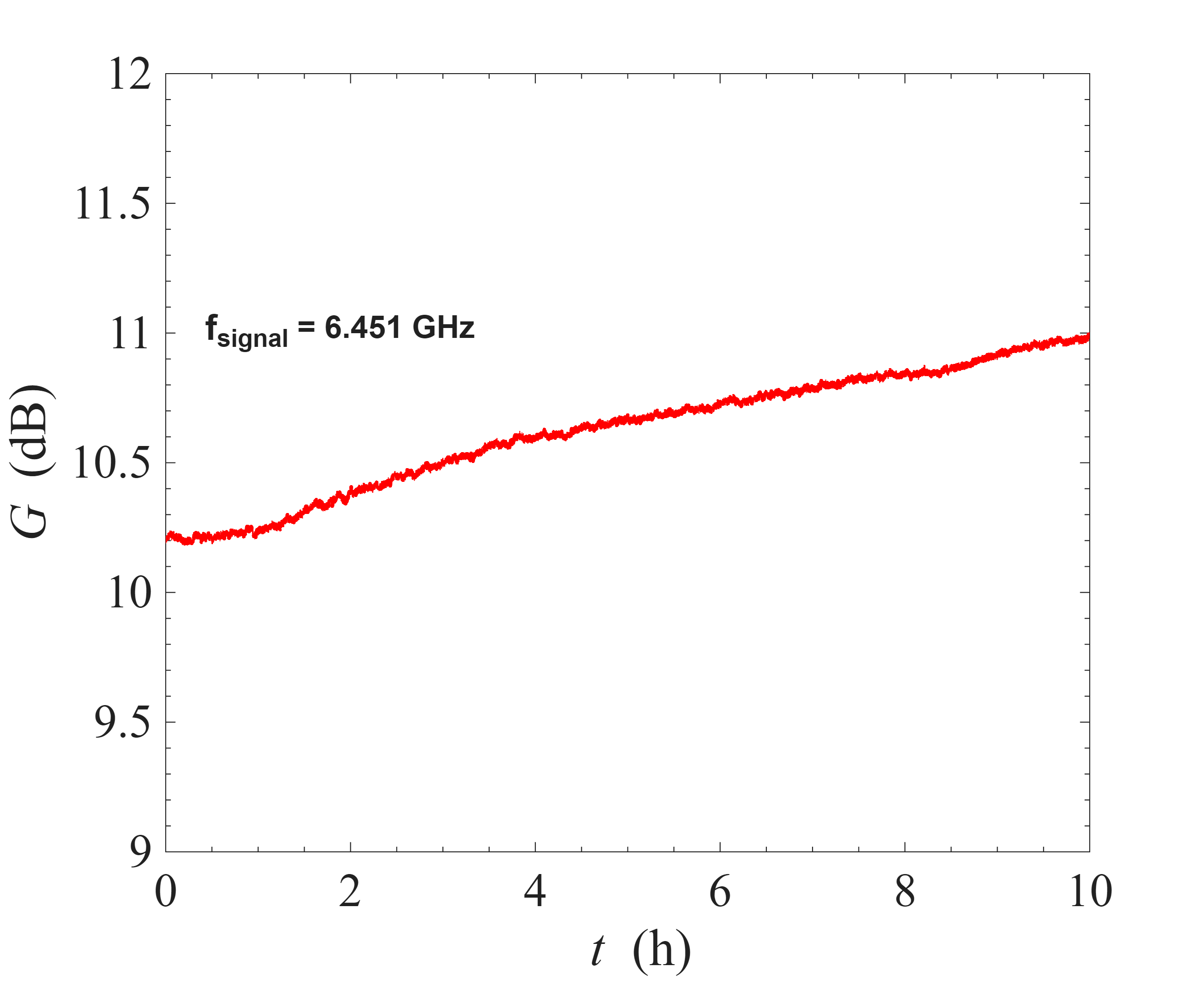}
\caption{\label{fig:13_4K_stability} Measurement of gain stability of a NbN resonator amplifier operated under non-degenerate pumping scheme at $3.2\,\mathrm{K}$. The signal frequency is fixed at $6.451\,\mathrm{GHz}$, where the gain is close to $10\,\mathrm{dB}$, and the signal gain measured continuously as a function of time for ten hours. The amount of drift in gain is approximately $0.075\,\mathrm{dB/h}$.}
\end{figure}
Another important measurement at $3.2\,\mathrm{K}$ is the characteristic stability of the ResPA under non-degenerate pumping scheme over timescale of several hours. We have fixed the signal frequency at $6.451\,\mathrm{GHz}$ where the gain was $\sim10\,\mathrm{dB}$ and monitored the gain continuously over ten hours. As shown in figure~\ref{fig:13_4K_stability}, the gain was stable and drifted by $0.75\,\mathrm{dB}$ over ten hours, translating to a slow drift of $0.075\,\mathrm{dB/h}$. We emphasise that this very small drift has been obtained even without any experimental effort at stabilising the measurement system in any aspect, e.g., the temperature of the ADR cryostat or the frequency of the signal generator for the pump tone. This gain drift is slightly higher than that obtained at $0.1\,\mathrm{K}$, but is still half as much compared to the same device under degenerate pumping, even at cryogenic temperatures of $0.1\,\mathrm{K}$ \citenumns{zhao2024_intrinsic_separation}. As discussed previously, this is likely a consequence of the non-degenerate pumping scheme placing the pump tones away from the bifurcation region where resonance properties change rapidly with frequency. 

These results at $3.2\,\mathrm{K}$ are highly promising, and suggest that ResPAs with non-degenerate pumping may be a highly important technology for high gain, low noise amplification at pulse tube cooler temperatures, and possibly as an alternative to HEMTs in narrow-band applications such as direct neutrino mass measurements \citenumns{Saakyan_2020,QTNM_collaboration_white_paper}.

\section{Discussion and Conclusion}
We have designed, realised, and measured an operating scheme for ResPAs by using non-degenerate pumping with two pump tones at different frequencies. As demonstrated by our measurements in the previous section, this scheme has several practical advantages over conventional degenerate pumping using a single pump tone. Firstly, the amplification band is continuous and it is not interrupted by a strong pump tone at its centre; secondly, the pump tones are placed several bandwidths away from the peak of the amplification profile, and can be removed more easily using low-Q filtering techniques to prevent saturation of downstream electronics; thirdly, the amplification profile under non-degenerate pumping is much more stable compared to the same device operated under degenerate pumping. Our measurements showed a factor of 4 improvement in stability in terms of gain drift per hour. Phase-sensitive amplification occurred at signal-idler degenerate frequency, and gain of $23\,\mathrm{dB}$ and squeezing ratio of $6\,\mathrm{dB}$ were measured.

The non-degenerate pumping technique for ResPAs is highly flexible: high gain can be achieved in a $\sim 4\,\mathrm{K}$ environment maintained by a pulse tube cooler, and in a cross-harmonic mode where the pumps are placed in adjacent resonance harmonics compared to the amplification band. $4\,\mathrm{K}$ operation is particularly exciting as it opens up the possibility of low-noise parametric amplification with low-cost cryogenic systems, and possibly as a HEMT replacement technology.

There are several future research directions that will be highly impactful in bringing the ResPA technology and the non-degenerate pumping technique closer to application. In general for ResPAs, improving the bandwidth to the order of a gigahertz will be particularly valuable in expanding the breadth of their potential applications, for example the search of axion dark matter \citenumns{Axion_2024}. Several techniques for expanding the bandwidth of ResPAs have been successfully demonstrated in the context of Josephson Parametric Amplifiers \citenumns{Roy_JPA_2015,Duan_JPA_2021,Naaman_2022,Kaufman_2023}. Next, direct measurement of the added noise from ResPAs will be important in demonstrating the noise advantage of parametric amplifiers. An exciting research direction would be to explore if and how the two pump tones in the non-degenerate pumping scheme can be generated using a single high-frequency pump tone modulated at a lower frequency, for example, through amplitude modulation. Successful research will reduce the cost-requirement of the non-degenerate pumping technique significantly. Finally, given the high peak gain measured in this study, a high level of squeezing could, in principle, be achieved in an optimised test system with well-matched components. Measuring the limit of squeezing ratio in such an optimised system would be impactful in determining the potential utility of ResPAs for squeezed amplification or to generate squeezed states for interferometric experiments. 

In summary, kinetic inductance ResPA is a highly flexible quantum technology that can operate in a wide range of temperatures and frequency bands, and can be fabricated with high yield in large numbers. Operating ResPAs under the non-degenerate pumping scheme yields important practical advantages, and enables phase-sensitive amplification. These developments demonstrate that ResPAs have the potential to be an important component in high-sensitivity quantum systems for low-noise parametric amplification, large-scale amplifier arrays, and squeezing-based interferometric systems.

\begin{acknowledgments}
The authors are grateful for funding from the UK Research and Innovation (UKRI) and the Science and Technology Facilities Council (STFC) through the Quantum Technologies for Fundamental Physics (QTFP) programme (Project Reference ST/T006307/2).
\end{acknowledgments}

\bibliographystyle{h-physrev}
\bibliography{library}
\end{document}